\documentclass[twocolumn,dvipsnames]{aastex62}
\usepackage[T1]{fontenc}
\usepackage[utf8]{inputenc}
\usepackage[english]{babel}
\usepackage{amssymb}
\usepackage{amsmath}
\usepackage{mathtools}
\usepackage{textgreek}
\usepackage{hyperref}
\usepackage{bookmark}
\usepackage{graphicx}
\usepackage{natbib}
\usepackage[version=4]{mhchem}

\renewcommand\familydefault{\rmdefault}
\DeclareMathAlphabet{\mathup}{OT1}{\familydefault}{m}{n}

\newcommand\tsup[1]{\textsuperscript{#1}}

\newcommand\supn[1]{\tsup{\ensuremath{#1}}}

\makeatletter
\newcommand\DeclareUnit[2]{%
    \@namedef{#1}{\@ifnextchar[{\csname @with@#1\endcsname}{\csname @without@#1\endcsname}}%
    \@namedef{@with@#1}[##1]{\text{#2\supn{##1}}}%
    \@namedef{@without@#1}{\text{#2}}%
}%
\makeatother

\DeclareUnit{km}{\text{km}}
\DeclareUnit{cm}{\text{cm}}
\DeclareUnit{mm}{\text{mm}}
\DeclareUnit{um}{\text{\textmu m}}
\DeclareUnit{s}{\text{s}}
\DeclareUnit{kg}{\text{kg}}
\DeclareUnit{g}{\text{g}}
\DeclareUnit{pc}{\text{pc}}
\DeclareUnit{au}{\text{au}}
\DeclareUnit{K}{\text{K}}
\DeclareUnit{Myr}{\text{Myr}}
\DeclareUnit{yr}{\text{yr}}

\newcommand\MSun{\ensuremath{M_\Sun}}

\newcommand\diff{\mathop{}\!\mathup{d}}

\newlength{\myarrow}
\newcommand*{\myrightarrow}[1]{\;\xrightarrow{\mathmakebox[\myarrow]{#1}}\;}

\newcommand\firstrevision[1]{#1}
\newcommand\secondrevision[1]{\textbf{#1}}

\begin{document}
\title{Chemistry Along Accretion Streams in a Viscously-Evolving Protoplanetary Disk}

\author[0000-0002-3286-3543]{Ellen M. Price}
\affiliation{Center for Astrophysics $|$ Harvard \& Smithsonian, 60 Garden St., Cambridge, MA 02138}

\author{L. Ilsedore Cleeves}
\affiliation{University of Virginia, Department of Astronomy, 530 McCormick Rd., Charlottesville, VA 22904}

\author{Karin I. \"Oberg}
\affiliation{Center for Astrophysics $|$ Harvard \& Smithsonian, 60 Garden St., Cambridge, MA 02138}

\shorttitle{Chemistry Along Accretion Streams in Protoplanetary Disks}
\shortauthors{Price, Cleeves, \& \"Oberg}

\begin{abstract}
The composition of a protoplanetary disk is set by a combination of interstellar inheritance and gas and grain surface chemical reactions within the disk. The survival of inherited molecules, as well as the disk \textit{in situ} chemistry depends on the local temperature, density and irradiation environment, which can change over time due to stellar and disk evolution, as well as transport in the disk. We address one aspect of this coupling between the physical and chemical evolution in disks by following accretion streamlines of gas and small grains in the disk midplane, while simultaneously taking the evolving star into account. This approach is computationally efficient and enables us to take into account changing physical conditions without reducing the chemical network. We find that many species are enhanced in the inner disk \firstrevision{midplane} in the dynamic model due to inward transport of cosmic-ray driven chemical products, resulting in, e.g., orders-of magnitude hydrocarbon enhancements at 1~\au, compared to a static disk. For several other chemical families, there is no difference between the static and dynamic models, indicative of a robust chemical reset, while yet others show  differences between static and dynamic models that depend on complex interactions between physics and chemistry during the inward track. The importance of coupling dynamics and chemistry when modeling the chemical evolution of protoplanetary disks is thus \firstrevision{depends} on what chemistry is of interest.
\end{abstract}


\section{Introduction}

% Why do we care about disk composition and chemistry?
Planets form in the dust- and gas-rich disks around young stars, i.e., in protoplanetary disks. The chemical composition of the disk directly impacts the compositions of forming planets and planetesimals. Rocky planets, ice giants, and comets all assemble mainly from disk solids, and their composition depends on the refractory and volatile content of dust grains. By contrast, gas giants obtain much of their mass directly from the disk gas, and their composition therefore depends on both the disk's gas and dust chemical compositions \citep[e.g.,][]{Mizuno1980PTP,BodenheimerPollack1986Icarus,Oberg+2011ApJL,Kokubo+2012PTEP,Cridland+2016MNRAS,Cridland+2017MNRAS}.

% Disk chemistry
%The radial location of forming planetesimals determines their compositions, because the 

% LIC: the last paragraph makes the transition to planet composition comes from disk composition, so disk conposition matters. Now we can jump to disk composition more directly:
The chemistry of the disk spatially varies due to gradients in radiation fields, temperature and density structures, and cosmic ray attenuation. These effects act to produce a radially-changing disk molecular composition \citep[e.g.,][]{Aikawa+1997ApJL,Willacy+2000ApJ,Bergin+2007book}. Moreover, at the relevant pressures and densities, much of the disk does not reach local \firstrevision{steady state} within the few million year lifetime of the gas disk \firstrevision{\citep[e.g.,][]{Aikawa+1998Faraday}}. Chemical processing of the material can thus have a significant effect on the planetesimals' compositions \citep{Eistrup+2016A&A}. Accurately predicting the time-evolving chemical history of a disk and comparing with observations of disk gas are therefore key to understanding what compositions planets can potentially acquire.

Observations of molecules in disks are limited \citep{mcguire2018} due to their low masses, relatively cool temperatures, and small angular extents. The inner disk ($R \lesssim 10~\au$) is observed to have gas-phase molecules like \ce{H2O}, \ce{CO}, nitrogen-bearing species, and hydrocarbons. \textit{Spitzer} observations of the upper disk layers in the inner disk suggest a partial ``reset-like'' chemistry \citep{Pontoppidan+2014} due to the extreme densities, temperatures, and radiation fields close to the star. As shown in this study, this chemistry is expected to be modified by continuous inward transport of molecules formed in the outer parts of the disk, resulting in deviations from steady-state chemistry.

The outer disk ($R \gtrsim 10~\au$), on the other hand, is characterized by sequential freeze-out of abundant volatiles \citep{HenningSemenov2013ChemRev} and a slower chemical conversion that may preserve more of the disk's initial composition, as set by the molecular cloud. In this region, the relative importance of inheritance and \textit{in situ} chemical processing depends on the vertical location within the disk. For \ce{H2O} ice in the midplane, models demonstrate that the outer disk likely exhibits an ``inheritance-like'' chemistry, where the timescales for chemical evolution can be longer than the disk lifetime, thus processing little of the ice \citep{SemenovWiebe2011ApJS,Cleeves+2014Science}. For cyanide species, observations are more consistent with chemical reprocessing \citep{Oberg+2015Nature}. 

These different results can, in part, be explained by strong vertical gradients in physical conditions in the outer disk. The uppermost, surface layer experiences strong irradiation, so relatively fast photochemistry dominates. Beneath the surface layer is a warm molecular layer where ion--molecule chemistry chemistry dominates; \ce{CO} is present but \ce{H2O} is frozen out onto grains. Finally, close to the midplane, molecules freeze out onto grains and slow grain surface reactions dominate the chemistry \citep{HenningSemenov2013ChemRev}.

% Disk dynamics
Modeling disk chemistry is complicated by the fact that disks and their host stars are dynamic --- densities, temperatures and radiation fields all evolve with time. Theory and observations show that protoplanetary disks are actively accreting \citep[e.g.,][]{LyndenBellPringle1974MNRAS,Gullbring+1998ApJ}, with some material transferred onto the star and some material moved outward, conserving angular momentum. Other dynamical processes that may be relevant include vertical mixing, turbulence, and grain drift and settling \citep{Whipple1973NASA,Weidenshilling1977MNRAS,Morfill1983Icarus,WeidenschillingCuzzi1993PRPL,Hanner1999SSR,BockeleeMorvan+2002A&A,Willacy+2006ApJ,Semenov+2006ApJL,Cridland+2017bMNRAS,Bacciotti+2018ApJ}. In addition to material physically moving in the disk, the star itself introduces an added complication since it, too, is evolving in time \citep[e.g.,][]{Siess+1997A&A}, changing the incident radiation field and temperature profile of the disk.

% Now get the reader excited for our approach!
%We can understand why dynamics should be important for chemistry in disks by considering that the timescale for radial flow due to viscosity is $t_\nu \sim R^2 / \nu$ \citep{ClarkeCarswell2007book}, where $\nu$, the kinematic viscosity, is assumed to be constant with radius (an assumption we will revise shortly). Take $\nu \sim 10^{13}~\cm[2]~\s[-1]$, an order-of-magnitude estimate based on the $\alpha$-disk model at $1~\au$ and $100~\K$. Under these assumptions, a parcel of gas at $1~\au$ would take of order $1~\Myr$ to reach the central star. Relevant chemical timescales span a range of magnitudes \citep{SemenovWiebe2011ApJS}. While it is impossible to determine \textit{a priori} which species' abundances may be altered by the introduction of dynamics, overlap between dynamical and chemical timescales suggests that, for some species, it may have a large effect.

The most common astrochemical treatment of a viscously-evolving disk is to fix the physical conditions to their local values at a given time and allow the chemistry to evolve at these fixed conditions. A handful of models have explored the coupling of some dynamical processes and chemistry. Early examples of coupling chemistry with accretion flows include \citet{Bauer+1997A&A} (and subsequent papers) and \citet{Aikawa+1999ApJ}. \citet{Nomura+2009A&A} followed the chemistry along simple streamlines in an $\alpha$-disk model and assumed a constant accretion rate. \citet{Heinzeller+2011ApJ} used the same physical model with the addition of X-ray heating and investigated the effects of radial accretion, turbulent mixing, and disk winds. Including accretion in the model caused significant changes in the chemical composition of the disk midplane. \citet{Cridland+2016MNRAS} and \citet{Cridland+2017MNRAS} used the \citet{Chambers2009ApJ} analytic model of viscous disk accretion to investigate the relationship between disk chemistry and planetary atmospheres, finding that the location and time at which the planet atmosphere accretes its atmosphere strongly affects its composition.

We build on these previous efforts and follow local physical conditions in accretion streams of gas and small, presumably coupled, grains in the disk midplane. We choose to focus on the midplane in this work because the gas and ice mantles on grains in the midplane eventually become available for incorporation into forming planetesimals, and because it allows some simplifying assumptions to be made regarding radiation fields and accretion stream geometries. This paper is structured as follows. In Section~\ref{sec:methods}, we introduce the method we use to trace both chemistry and physical conditions as functions of time. In Section~\ref{sec:results}, we present the results of our model. We discuss and conclude our analysis in Sections~\ref{sec:discussion} and \ref{sec:conclusions}, respectively.

\section{Methods}
\label{sec:methods}

\firstrevision{Our goal is to compute the time-dependent tracks that gas follows through a vicously evolving disk around an evolving pre-main sequence star and to solve for the chemical abundances along these tracks. Note that the tracks are treated as completely independent from one another, i.e., are not mixed, greatly simplifying the computation. Additionally, we treat the dust as well-coupled to the gas and do not address dust grain evolution, which may be chemically important. The main role of the dust is to provide surface area for chemical reactions. Dust growth and fragmentation fundamentally change the surface area to volume ratio of dust, however the dynamical evolution is more complex than can be treated in the present prescription, which we will address in future work.}

\begin{figure}
    \centering
    \includegraphics[width=0.8\linewidth]{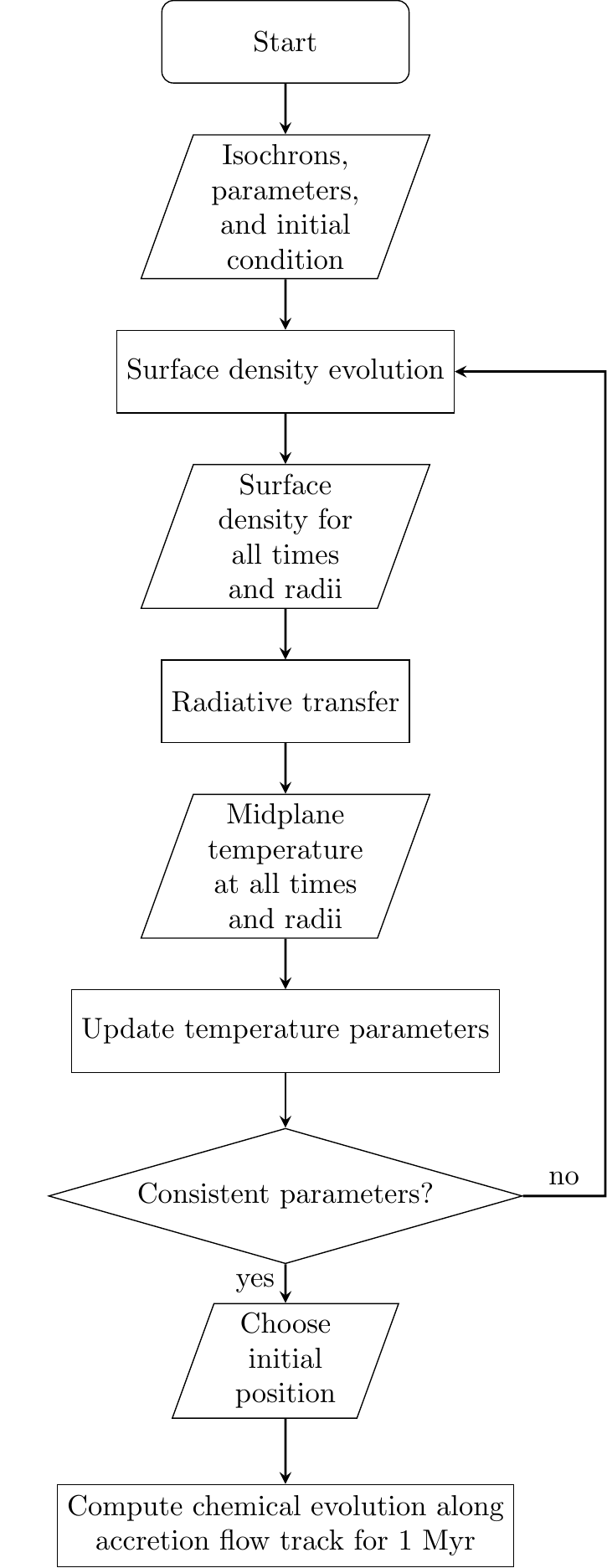}
    \caption{Outline of the method presented in Section~\ref{sec:methods}. In this diagram, rectangles represent processes, trapezoids represent inputs and outputs, and diamonds represent decisions. We begin by making an initial guess of parameters $T_0$, $\psi$, $\omega$, $\alpha_0$, $\alpha_1$, and $\beta_1$ (see Equation~\ref{eqn:temp-evolution}) and iterate between solving Equation~\ref{eqn:surface-density-evolution} and using radiative transfer until a consistent set of parameters is found. At that point, we can choose any initial point and solve for a gas parcel's trajectory through the disk from Equation~\ref{eqn:tracks}. Finally, the chemical evolution is solved in a postprocessing step.}
    \label{fig:flowchart}
\end{figure}

One of the primary challenges to overcome in this method is that the surface density evolution and temperature structure are interdependent. To solve for both self-consistently, we use an iterative procedure, outlined in Figure~\ref{fig:flowchart}.

% TODO: Ilse thinks we need another step in the flowchart chain, right before the last step. Given the methods section as currently written, what needs to be included?

\subsection{Accretion disk model}
\label{sec:physical-model}
We construct the physical disk model using the $\alpha$-disk framework \citep{ShakuraSunyaev1973A&A}. We work in a cylindrical coordinate system parametrized by $\left(R, \phi, z\right)$; here, $R$ is the radial coordinate (distance from the star in the $x$-$y$ plane), $\phi$ is the azimuthal angle, and $z$ is the vertical coordinate (height above the midplane).

The viscosity of an $\alpha$-disk is given by
\begin{equation}
    \nu = \alpha c_s h = \alpha c_s^2 / \Omega_{\mathrm{Kep}}
\end{equation}
 where $\Omega_{\mathrm{Kep}} \equiv \sqrt{G M_\star / R^3}$ is the Keplerian angular velocity, $c_s$ is the local sound speed, $h$ is the local disk scale height \firstrevision{determined by the midplane temperature}, and $M_\star$ is the stellar mass; we assume that the disk mass is negligible compared to $M_\star$. $\alpha$ is a small dimensionless parameter with typical values $10^{-2}$ to $10^{-4}$; this range is supported observationally by, e.g., \citet{Flaherty+2018ApJ}, which found $\alpha < 0.007$ for the TW Hya disk. The sound speed is given by
\begin{equation}
	c_s^2 = \frac{k_B T}{\mu m_p},
\end{equation}
where $\mu$ is the mean molecular weight of the gas, $k_B$ is the Boltzmann constant, and $T$ is the temperature of the gas and dust; we take the gas temperature equal to the dust temperature, which is reasonable in the disk midplane where the dust and gas are well-coupled through collisions \citep{HenningSemenov2013ChemRev}. This temperature is computed and subsequently parametrized as a function of time and radius (see Section~\ref{sec:stellar-evolution}).

We begin with the general surface density evolution equation \citep[e.g.,][]{Pringle1981ARAA,ClarkeCarswell2007book,Armitage2010book},
\begin{equation}
	\frac{\partial \Sigma}{\partial t} - \frac{3}{R} \frac{\partial}{\partial R} \left[R^{1/2} \frac{\partial}{\partial R} \left(\nu \Sigma R^{1/2}\right)\right] = 0.
    \label{eqn:surface-density-evolution}
\end{equation}
This equation can be derived from the Navier-Stokes and mass continuity equations for a fluid, as shown in \citet{ClarkeCarswell2007book}. Because of the flexible form of the temperature we choose (see Section~\ref{sec:stellar-evolution}), there exists no easily-found analytic solution for the surface density. Thus, we solve the equation numerically, using a simple finite difference scheme with second-order accurate spatial derivatives and a Crank-Nicoloson timestepping scheme. We implement this method using PETSc \citep{petsc-user-ref,petsc-efficient,tspaper}.

Our initial surface density profile is informed by observations of disks, so we choose a form similar to
\begin{equation}
	\Sigma\!\left(t = 0, R\right) \propto \left(\frac{R}{R_1}\right)^{-\gamma} \exp\!\left[-\left(\frac{R}{R_1}\right)^{2-\gamma}\right]
    \label{eqn:self-similar}
\end{equation}
in the notation of, e.g., \citet{Andrews+2012ApJ}. However, for $\gamma = 1$, a reasonable value based on observational fitting, this initial condition would approach infinity as $R$ approaches zero. This presents a computational challenge, because the value of $\Sigma$ cannot simply be fixed to a value at small radii due to disk evolution, yet an infinite value at $R = 0$ is both unphysical and difficult to handle numerically. We circumvent this problem by introducing a sharp exponential taper at finite radius, given by
\begin{equation}
	f\!\left(R\right) = \begin{cases}
    	\exp\!\left[-\left(\frac{R - R_3}{R_4}\right)^{2\xi}\right], & R < R_3 \\
        1, & R \geq R_3
    \end{cases}
    \label{eqn:fudge-factor}
\end{equation}
with shape parameters $R_3$, $R_4$, and $\xi > 0$ that may be chosen freely. This function decays faster than Equation~\ref{eqn:self-similar} blows up\footnote{At first glance, it may appear that imposing zero surface density at finite radius inhibits accretion. However, if we examine the functional form of the accretion rate, $\dot{M} = -2\pi R \Sigma u_R$ \citep{Pringle1981ARAA}, we see from Equation~\ref{eqn:tracks}, which gives $u_R$, that the factors of $\Sigma$ cancel; so the accretion rate can still be finite when $\Sigma$ approaches zero.}, so it is effective in producing the desired Dirichlet boundary condition\footnote{A boundary condition in which the value of the unknown function is fixed to a given value at the boundaries only; in this case, the value zero is enforced.} at small $R$. Because Equation~\ref{eqn:self-similar} decays exponentially at large $R$, we do not need to introduce additional factors to produce a Dirichlet boundary condition at $R / R_1 \gg 1$. In practice, then, we multiply Equation~\ref{eqn:self-similar} by Equation~\ref{eqn:fudge-factor} and then normalize to a chosen value $\Sigma_2 = \Sigma\!\left(t = 0, R_2\right)$, where $R_2$ is a chosen normalization radius. Parameters relevant to this model are listed in Table~\ref{tbl:params}.

\begin{deluxetable}{lcc}
\tablecaption{Fiducial model parameters. \label{tbl:params}}
\tablehead{%
\colhead{Parameter} & \colhead{Symbol} & \colhead{Value} %
}
\startdata
Stellar mass & $M_\star$ & $1 \MSun$ \\
Viscosity parameter & $\alpha$ & $10^{-3}$ \\
Mean molecular weight & $\mu$ & $2.35$ \\
Normalization radius & $R_0$ & $1~\au$ \\
Shape parameter & $R_1$ & $40~\au$ \\
$\Sigma$ normalization radius & $R_2$ & $10~\au$ \\
Shape parameter & $R_3$ & $0.3~\au$ \\
Shape parameter & $R_4$ & $0.1~\au$ \\
Value of $\Sigma$ at $R_2$ & $\Sigma_2$ & $100~\g~\cm[-3]$ \\
Exponential taper exponent & $\xi$ & 4 \\
Gas-to-dust ratio & & $100$ \\
Cosmic ray rate & $\zeta_\mathrm{CR}$ & $10^{-18}~\s[-1]$
\enddata
\end{deluxetable}

\begin{figure*}
	\centering
    \includegraphics[width=\linewidth]{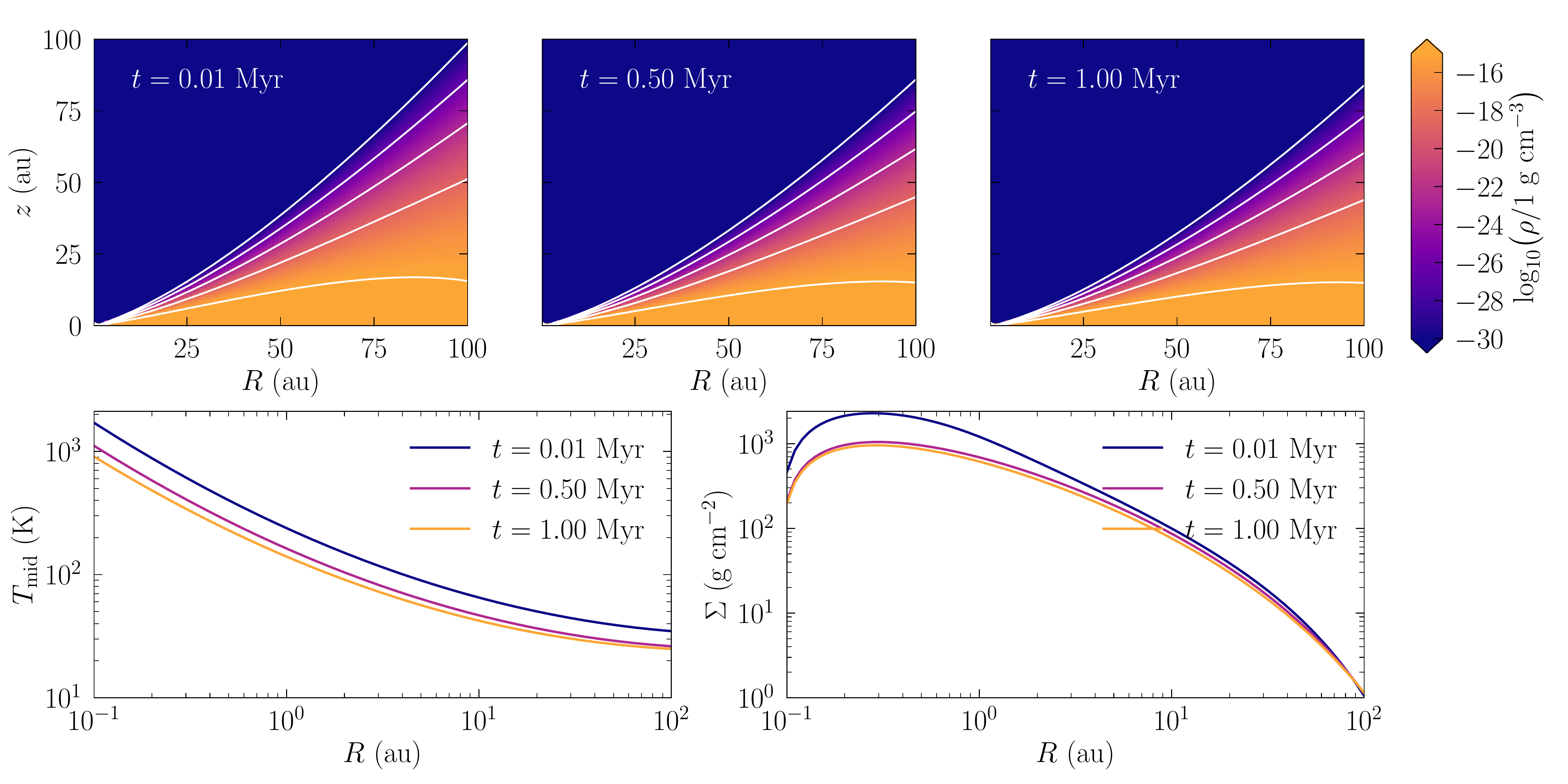}
    \caption{Disk density, midplane temperature, and surface density sampled at different times, computed by solving the equations in Section~\ref{sec:physical-model}. The top three panels show the mass density $\rho$ as a function of radius, $s$, height, $z$, and time. The bottom panels show the midplane temperature and surface density, both as functions of radius and time.}
\end{figure*}

\subsection{Disk temperature and density structures}
\label{sec:stellar-evolution} 

As mentioned previously, we must assume a temperature structure to solve for the surface density, ultimately resulting in a new temperature structure. We use the \citet{Siess+2000A&A} isochrons for a $M_\star = 1 \MSun$, $Z = 0.02$ star to obtain the stellar radius and effective temperature on a linearly-spaced grid of given ages. We combine the dust density, assuming azimuthal symmetry, with the stellar parameters and compute the resulting dust temperature structure using RADMC-3D \citep{radmc3d}; at the end of the Monte Carlo simulation, the effects of accretion heating are added in flux space, so
\begin{equation}
    T^4_\mathrm{final} = T^4_\mathrm{RADMC} + T^4_\mathrm{accretion}
\end{equation}
\firstrevision{and
\begin{equation}
    T_\mathrm{accretion}^4 = \frac{G M_\star \dot{M}}{8 \pi \sigma R^3}
\end{equation}
\citep[e.g.][]{Hartmann2009book}}. We iterate this procedure, fitting each time for the unknown parameters in our temperature model until the solutions converge to a relative precision of $5\%$.

We assume a temperature function --- which enters into Equation~\ref{eqn:surface-density-evolution} through the viscosity term --- that takes the flexible form 
\begin{equation}
 	T\!\left(t, R\right) = T_0 \left(e^{-\psi t / t_0} + \omega\right) e^{\beta_0 \log x + \beta_1 \log^2 x},
 	\label{eqn:temp-evolution}
\end{equation}
where $x = R / R_0$ is a dimensionless radius; $\beta_0 = \alpha_0 t / t_0 + \alpha_1$; and $T_0$, $\psi$, $\omega$, $\alpha_0$, $\alpha_1$, and $\beta_1$ are parameters to be determined. We find that this large number of parameters is necessary to capture, with reasonable accuracy, the full radial- and time-dependent behavior of the disk temperature structure. The approximate final parameters we derive for this model are listed in Table~\ref{tbl:tempparams}. \firstrevision{We do not use the raw temperatures from RADMC-3D because derivatives of the temperature function are needed for the time evolution of the surface density, and using the output without fitting would introduce unnecessary noise.}

\begin{deluxetable}{cccc}
\tablecaption{Fiducial model temperature parameters from iterative fitting procedure. \label{tbl:tempparams}}
\tablehead{%
\colhead{Symbol} & \colhead{Value} & \colhead{Symbol} & \colhead{Value} %
}
\startdata
$T_0$ & $110$~\protect\K & $\alpha_0$ & 0.044 \\
$\psi$ & 2.5 & $\alpha_1$ & -0.71 \\
$\omega$ & 1.2 & $\beta_1$ & 0.063
\enddata
\end{deluxetable}

\begin{figure}
	\centering
    \includegraphics[width=\linewidth]{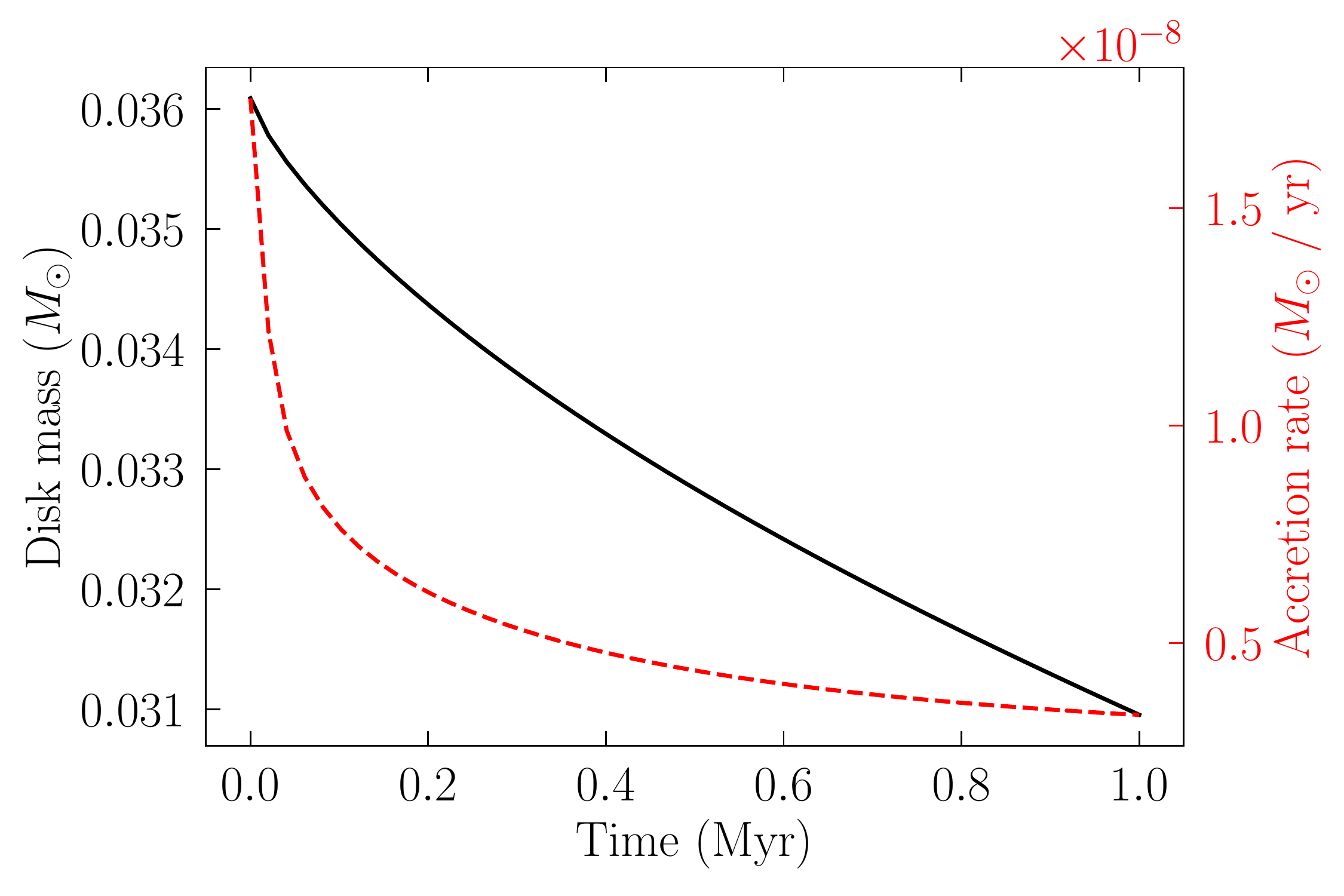}
    \caption{Total disk mass (black, solid line) and accretion rate (red, dashed line) as a function of time.}
    \label{fig:mass}
\end{figure}

Once the temperature structure is known, the disk structure is fully determined at all times and radii. This information allows us to compute, for example, the disk mass and accretion rate as functions of time. These quantities are shown in Figure~\ref{fig:mass}, from which we confirm that the accretion rate is reasonable.

We solve Equation~\ref{eqn:surface-density-evolution} for a given set of temperature parameters to obtain the surface density at all radii and all times. We interpolate this function and obtain the gas density by assuming that it has a vertical Gaussian profile, i.e.

\begin{equation}
	\rho\!\left(R, z\right) = \frac{\Sigma\left(R\right)}{\sqrt{2 \pi} h} \exp\!\left(-\frac{z^2}{2 h^2}\right).
\end{equation}
Further we adopt a gas-to-dust mass ratio of 100 at all radii, and assume only small micron-sized grains as prescribed in \citet{Fogel+2011ApJ}, such that the dust grains are well-coupled to the gas motion; we do not include larger grains here.

\subsection{Computing tracks}
\label{sec:computing-tracks}

\begin{figure*}
	\centering
    \includegraphics[width=\linewidth]{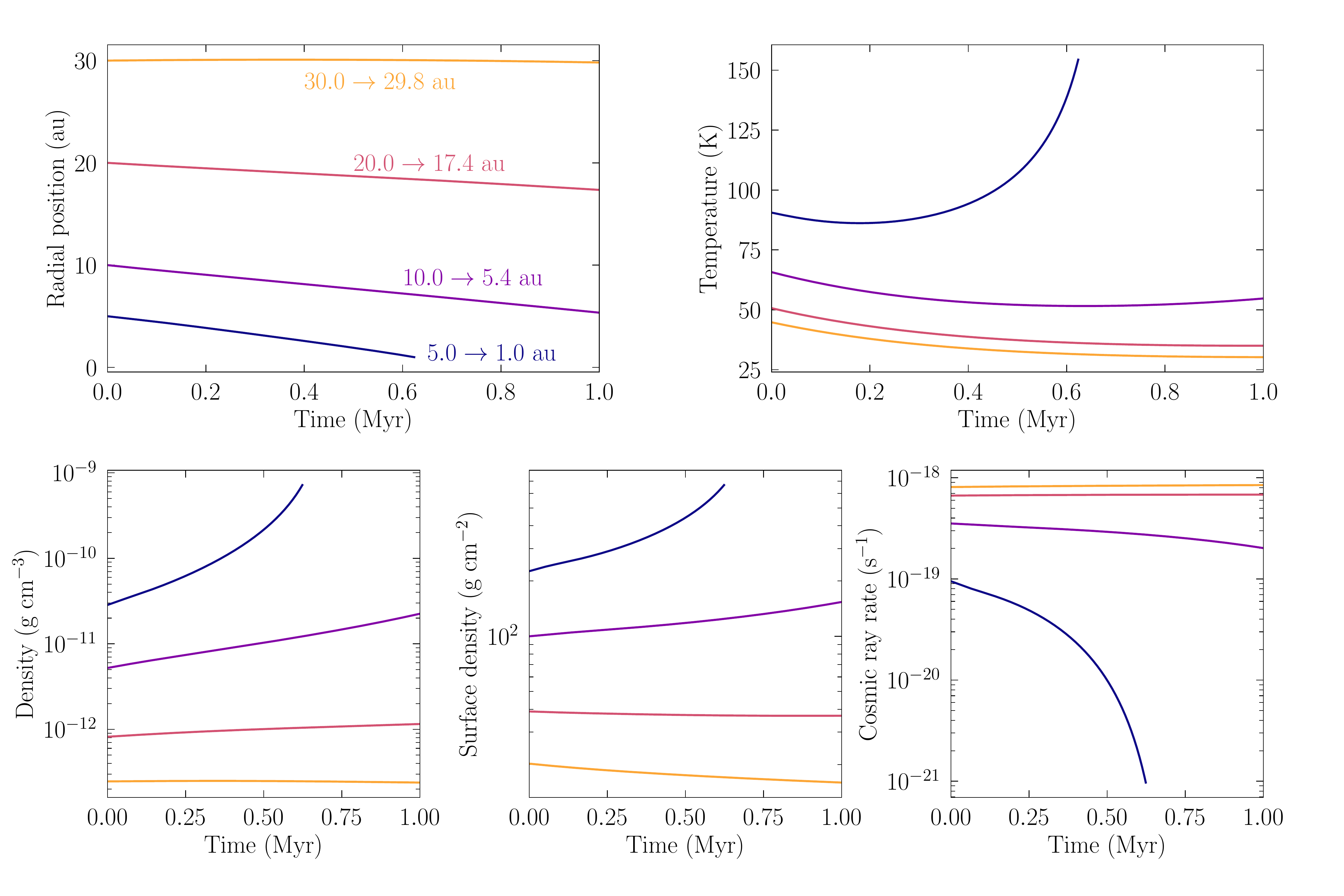}
    \caption{Change in physical variables (temperature, density, surface density, and cosmic ray rate) along tracks through the disk, as a function of starting radius. Note that density, surface density, and temperature changes with time are due to a combination of inward movement, which generally implies increasing density, surface density and temperature, an overall decreasing disk mass, and an evolving star, which decreases in luminosity with time. The strong decrease in cosmic ray ionization rate during the $5 \rightarrow 1$~\protect\au\ track is due to efficient attenuation of cosmic rays in the inner, high-surface density disk. Tracks are computed as described in Section~\ref{sec:computing-tracks}.}
    \label{fig:tracks}
\end{figure*}

For the physical model and stellar evolution described in the previous sections, we can compute self-consistent tracks of material through the disk following the velocity field implied by the evolving surface density profile. These tracks are given by the solution to the differential equation
\begin{equation}
    u_R = -\frac{3}{R^{1/2} \Sigma} \frac{\partial}{\partial R} \left(R^{1/2} \nu \Sigma\right)
    \label{eqn:tracks}
\end{equation}
\citep[e.g.,][]{ClarkeCarswell2007book}, \firstrevision{where $u_R$ is the velocity in the radial direction}. Once the solution for $\Sigma$ is known, this equation may be solved numerically. In Figure~\ref{fig:tracks} we show several representative solutions with the temperature, density, surface density, and cosmic ray rate evaluated along the midplane tracks. Note that for many of the tracks the changes in temperature, density, and cosmic ray ionization are small to moderate. The only track that experiences order of magnitude changes in any of these physical variables is the 5$\rightarrow$1~au track. 

We follow tracks in the disk midplane for $1~\Myr$, the typical lifetime of a protoplanetary disk \citep{Furlan+2009ApJ,Mamajek2009AIP}, with the exception of tracks that cross 1~\au, in which case we stop the track as it is crossing. We impose the radius restriction to avoid any effects from our inner boundary condition, which, as discussed above, was imposed to ensure that the solution is unique. Though dust growth will occur on similar time scales as those we consider, this effect is ignored in the present study to isolate the effects of gas dynamics.

\subsection{Disk chemistry}
\label{sec:disk-chemistry}

Our disk chemistry code builds on that of \citet{Fogel+2011ApJ} and \citet{Cleeves+2014ApJ}, which itself is a modified version of ALCHEMIC \citep{Semenov+2010AA}. The \citet{Fogel+2011ApJ} code calculates the chemical evolution in zones that are completely independent and stationary. We instead consider parcels that are independent but not stationary, following accretion tracks through the disk, as described above. Because the tracks do not cross, the chemical evolution can be treated as a postprocessing step once the surface density model is determined.

To compute the change in abundance of each chemical species as a function of time, we must account for the fact that both the number of each species \textit{and} the volume\footnote{The volume of the parcel is changing because, as the parcel approaches the star, the local density increases; as we are neither creating nor destroying matter, this directly corresponds to a volume decrease.} of the gas parcel are changing as functions of time, due to the chemistry and dynamics, respectively.

We denote the number density of a single species $i$ as $n_i \equiv N_i / V$, where $N_i$ is the number of species $i$ and $V$ is the volume of the gas parcel. Applying the quotient rule for derivatives to $n_i$, we find that
\begin{equation}
    \frac{\diff n_i}{\diff t} = \frac{1}{V^2} \left(V \frac{\diff N_i}{\diff t} - N_i \frac{\diff V}{\diff t}\right) = \frac{\diff n_i}{\diff t}\bigg|_V \!\!+ \frac{\diff n_i}{\diff t}\bigg|_{N_i}.
    \label{eqn:dndt}
\end{equation}
We can identify the first term as the rate of change of $n_i$ due to chemistry alone and the second term as the rate of change of $n_i$ due to dynamics alone.

The first term, the rate of change due to chemistry, is straightforward to compute given a chemical network defining all possible reactions. Consider two types of chemical reactions, those for which $n_i$ is a reactant (Equation~\ref{eqn:reactant}) and those for which $n_i$ is a product (Equation~\ref{eqn:product}):
\begin{align}
    n_i + n_j &\rightarrow \cdots \label{eqn:reactant} \\
    n_{j1} + n_{j2} &\rightarrow n_i + \cdots \label{eqn:product}
\end{align}
 Let $R_j$ be the rate of reaction $j$ for which $n_i$ is a reactant and let $P_j$ be the rate of reaction $j$ for which $n_i$ is a product. Then,
\begin{equation}
    \frac{\diff n_i}{\diff t}\bigg|_V = \sum\limits_j P_j n_{j1} n_{j2} - n_i \sum\limits_j R_j n_j.
    \label{eqn:dndt-constant-volume}
\end{equation}
To write the second term, the rate of change due to dynamics, in terms of quantities we know, we apply the chain rule for derivatives, where the total mass density $\rho \equiv M / V$ for an unchanging parcel mass $M$:
\begin{equation}
    \frac{\diff n_i}{\diff t}\bigg|_{N_i} \!\!= N_i \frac{\diff \!\left(\frac{1}{V}\right)}{\diff t} = N_i \frac{\diff \!\left(\frac{1}{V}\right)}{\diff \rho} \frac{\diff \rho}{\diff t} = \frac{N_i}{M} \frac{\diff \rho}{\diff t} = \frac{n_i}{\rho} \frac{\diff \rho}{\diff t}
    \label{eqn:dndt-constant-number}
\end{equation} 

Using Equation~\ref{eqn:dndt}, we construct a simultaneous system of coupled ordinary differential equations. We solve the system of equations using the \textsc{cvode} code from the Lawrence Livermore National Laboratory \textsc{sundials} package \citep{Hindmarsh+2005ACM}; this code was chosen because it is efficient, under current development, and easily accessed from C. The system is stiff (meaning that reactions occur on many different timescales), so we choose the backward differentiation formula (BDF) method with Newton iteration. The linear system is large and sparse, so we use MUMPS \citep{MUMPS:1,MUMPS:2} as a direct solver, through the PETSc interface. Our initial chemical abundances are given in Table~\ref{tbl:initial_conditions} and are characteristic of observed molecular cloud and protostellar envelope abundances; \firstrevision{the values are inspired by \citet{AikawaHerbst1999ApJ}, updated by \citet{Fogel+2011ApJ}, and further updated with ice abundances from observations of protostellar envelopes \citep{Boogert2015}}. The physical conditions are updated at each time step according to the surface density solution, which need only be computed once.

\begin{deluxetable}{lclc}
\tablecaption{Initial chemical abundance, which assumes inheritance from the molecular cloud stage. The species' abundances are given with respect to total hydrogen. \label{tbl:initial_conditions}}
\tablehead{%
Species & \colhead{Abundance} & Species & \colhead{Abundance}%
}
\startdata
\ce{H2} & $5.0 \times 10^{-1}$ & \ce{O} & $1.0 \times 10^{-8}$ \\
\ce{O2} & $1.0 \times 10^{-8}$ & \ce{He} & $1.4 \times 10^{-1}$ \\
\ce{N2} & $3.6 \times 10^{-5}$ & \ce{CN} & $6.0 \times 10^{-8}$ \\
\ce{H3^+} & $1.0 \times 10^{-8}$ & \ce{S^+} & $1.0 \times 10^{-11}$ \\
\ce{Si^+} & $1.0 \times 10^{-11}$ & \ce{Mg^+} & $1.0 \times 10^{-11}$ \\
\ce{Fe^+} & $1.0 \times 10^{-11}$ & \ce{CO} & $1.0 \times 10^{-4}$ \\
\ce{C} & $5.0 \times 10^{-9}$ & \ce{NH3} & $8.0 \times 10^{-8}$ \\
\ce{HCN} & $2.0 \times 10^{-8}$ & \ce{C^+} & $1.0 \times 10^{-10}$ \\
\ce{HCO^+} & $9.0 \times 10^{-9}$ & \ce{C2H} & $8.0 \times 10^{-9}$ \\
\ce{H2O}(gr) & $8.4 \times 10^{-5}$ & \ce{CO}(gr) & $1.5 \times 10^{-5}$ \\
\ce{CO2}(gr) & $2.4 \times 10^{-5}$ & \ce{CH3OH}(gr) & $4.2 \times 10^{-6}$ \\
\ce{NH3}(gr) & $3.4 \times 10^{-6}$ & \ce{CH4}(gr) & $2.5 \times 10^{-6}$ \\
GRAIN0 & $6.0 \times 10^{-12}$ & GRAIN & $6.0 \times 10^{-12}$
\enddata
\end{deluxetable}

\section{Results}
\label{sec:results}

To explore the importance of gas accretion dynamics for the chemical evolution of a disk, we run three simulations for a given starting radius: one with the full dynamical treatment (the dynamic model), one with a fixed position at the initial radius of the track (the initial point evolution model), and one with a fixed position at the end radius of the track (the final point evolution model). Though the radius $R$ remains fixed for these static models, the local physical conditions are allowed to vary as the disk evolves. All simulations are run for 1~\Myr, except when a track reaches 1~\au\ before that time has elapsed.

\subsection{Effect of adding dynamics}

\begin{figure*}[ht!]
    \centering
    \includegraphics[width=\linewidth]{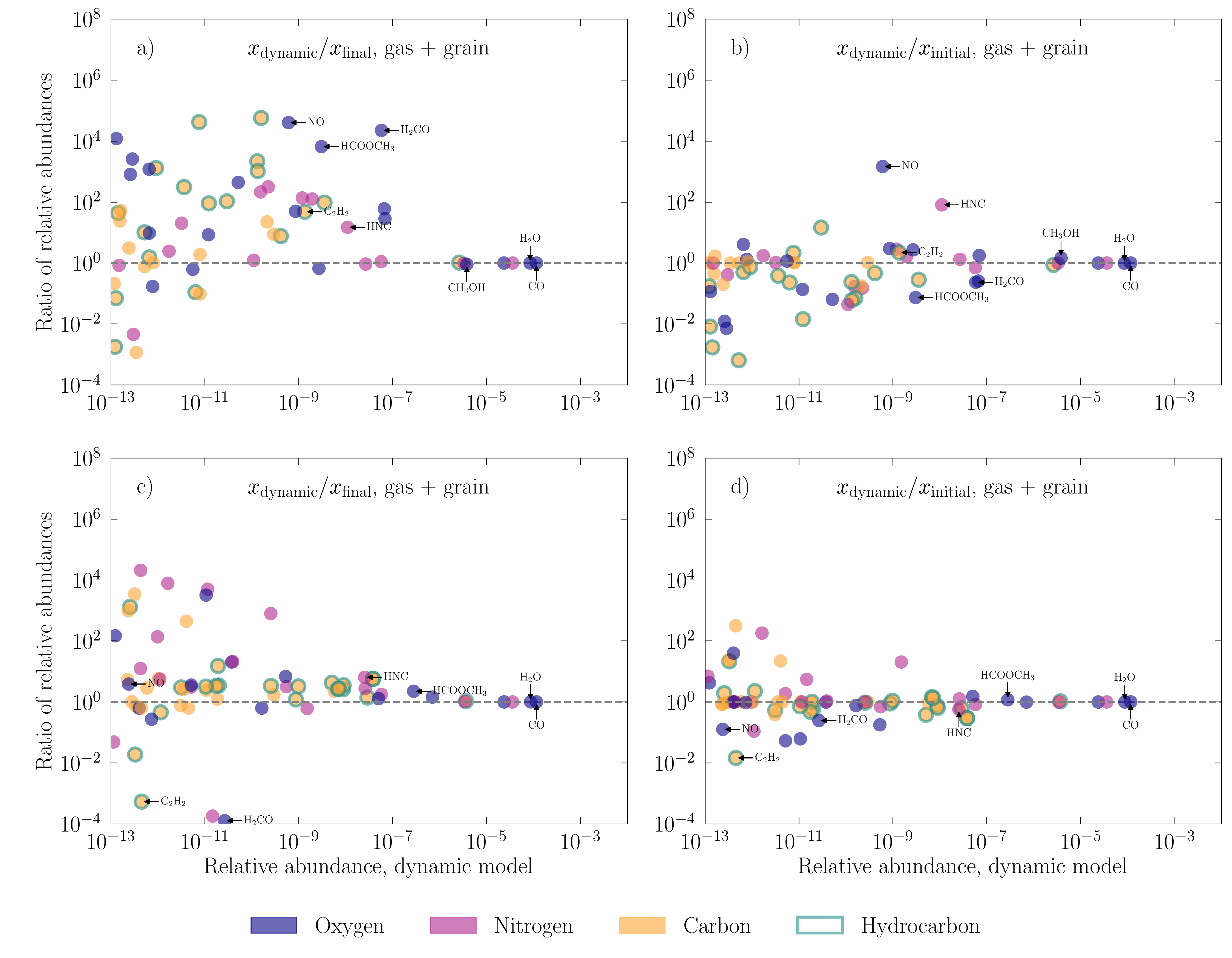}
    \caption{For dynamic tracks that evolve from 5~\protect\au\ (top row) and 10~\protect\au\ (bottom row), we show the relative abundances (gas + grain) of various species in our dynamical model compared to the final point evolution (left column) and to the initial point evolution (right column). Species are color-coded by the heaviest atom in the species; \firstrevision{for example, the heaviest atom in \ce{CO2} would be oxygen, and the heaviest atom in HNC would be nitrogen}. Interesting chemical families have color-coded borders as indicated in the lower key. The dashed horizontal line indicates where the abundances would be identical in the dynamic and static models. Any species that appears above the line is overproduced when dynamics are included. Molecules that are discussed in the text are labelled.}
    \label{fig:results2}
\end{figure*}

In Figures~\ref{fig:results2}a and \ref{fig:results2}b, we compare the \textit{total} (that is, gas and solid phases combined) final relative abundances\footnote{Throughout this section and others, we will refer to the ``relative abundance'' of a species. Generally, one normalizes the absolute abundance by $[\ce{H}] + 2 [\ce{H2}]$. However, we find that there may be substantial amount of hydrogen in less abundant species, which can introduce a systematic error in the relative abundances. To avoid this problem, we total the amount of hydrogen across all species and normalize by this quantity.} of the $5 \rightarrow 1$~\au\ model to those of the two static point models at 5~\au\ and 1~\au. By comparing the total quantity, we remove the effects of sublimation, i.e., snowline crossings, enabling us to isolate overall compositional changes. In each case, the ratio of abundances between the dynamic and static models are plotted against the abundance with respect to hydrogen in the dynamic model, all at the end of the simulation. 

Chemical families of interest have been highlighted with colored outlines, and the inner color of each point corresponds to the heaviest atom in the molecule. Figure \ref{fig:results2} shows that most species are enhanced in the dynamic model compared to the final point model, often by orders of magnitude. The notable exceptions are the handful of highly abundant species at the model inception, such as H$_2$O, CO and CH$_3$OH.  There are $\sim$20 substantially-enhanced ($x_\mathrm{dynamic} / x_\mathrm{static} > 10$) and abundant ($x_\mathrm{dynamic} > 10^{-10}$) species at the end of the $5 \rightarrow 1$~\au\ dynamic model track as compared to the static point model run at 1~\au, and these are listed in Table~\ref{tbl:total-enhanced-5au}.

Most of the enhanced species are carbon chains, more saturated hydrocarbons, and nitriles. There are also a few complex organic molecules that are ehanced, such as \ce{CH3OCH3} and \ce{HCOOCH3}. Only two species, \ce{HNC} and \ce{NO} (see Table~\ref{tbl:total-enhanced-5au}), are more than one order of magnitude enhanced when comparing the $5 \rightarrow 1$~\au\ dynamic track and the 5~\au\ (initial point) static model. This behavior indicates that much of the 1~\Myr\ chemical composition in the dynamic track is set by reactions close to the starting point of the track, which is then transported inwards. The depletion of many species compared to the initial point model (Figure \ref{fig:results2}) shows that the survival is not perfect, however, and both transport and local chemistry needs to be taken into account when modeling chemical abundances in the inner disk.

Figures~\ref{fig:results2}c and \ref{fig:results2}d show the analogous data for the track beginning at 10~\au. In contrast to the $5\rightarrow1$~au track, the $10\rightarrow5$~au track presents few species that are both abundant and substantially enhanced compared to the final point model. A quite small difference of 5~\au\ vs. 10~\au\ in starting radius thus result in a large difference when evaluating the role of dynamics in disks. Furthermore, there is not necessarily a consistent enhancement/depletion pattern between the two scenarios. Some species, such as \ce{C2H2}, which were enhanced in the 5~\au\ dynamic model compared to the final point static model, are now \textit{depleted} when adding dynamics to the 10~\au\ model (Figure~\ref{fig:results2}c). %When comparing to the initial point chemistry, \ce{C2H2} is depleted by a factor of $> 10$, and when compared to the final point model by a factor of $> 10^3$. 
%Some species, such as \ce{C2H2}, which were enhanced in the 5~\au\ dynamic model compared to the static model are depleted when adding dynamics to the 10~\au\ model --- in this case, by a factor $> 10$, is depleted in the 10~\au\ dynamic model compared to the 10~\au\ final point model by a factor $> 10^3$. 

For tracks starting at radii $R \geq 20~\au$,  we see little to no chemical difference between the initial, final, and dynamic models. This can be understood when considering that the physical conditions do not change significantly along these tracks, and, where there are changes, they tend to be towards lower temperatures and densities, which typically decrease the chemical reaction rates.

In summary, adding dynamics in the form of inward accretion streams has a large effect on the chemical evolution of the innermost disk as exemplified by the 5 to 1~\au\ track, while the effects on the chemical evolution exterior to 5~\au\ are small. In all models, adding accretion dynamics does not affect the abundances of the initially most abundant molecules such as \ce{CO} and \ce{H2O}, which on our timescales are not substantially destroyed or enhanced. %The overall abundance of \ce{CO} does not change significantly in any of the static or dynamic models, and \ce{H2O} equilibrates to its new environment so quickly that its final abundance is almost identical to that of the static model at the endpoint of the track.

% CO probably under goes *lots* of creation and destruction however the overall change in abundance is likely quite small, so i tried to rephrase. also 

% XXX: what else would be good to include here?

\begin{deluxetable}{lcc}[h]
\tablecaption{Total (gas + grain) abundances of significantly enhanced, abundant species in the dynamic model compared to the two static point evolution models for the 5~\protect\au\ track. \label{tbl:total-enhanced-5au}}
\tablehead{%
& \colhead{Abundance w.r.t. hydrogen} & \colhead{Enhancement} %
}
\startdata
\multicolumn{3}{l}{\textit{Comparing to final point model}} \\
\hline
C${}_{2}$H${}_{2}$ & $1.3 \times 10^{-9}$ & $4.8 \times 10^{1}$\\
C${}_{2}$H${}_{2}$N & $2.2 \times 10^{-10}$ & $3.2 \times 10^{2}$\\
C${}_{2}$H${}_{3}$N & $1.5 \times 10^{-10}$ & $2.1 \times 10^{2}$\\
C${}_{2}$H${}_{4}$ & $3.5 \times 10^{-9}$ & $9.6 \times 10^{1}$\\
C${}_{3}$ & $2.1 \times 10^{-10}$ & $2.2 \times 10^{1}$\\
C${}_{3}$H${}_{2}$ & $1.3 \times 10^{-10}$ & $2.2 \times 10^{3}$\\
C${}_{3}$H${}_{3}$ & $1.6 \times 10^{-10}$ & $5.8 \times 10^{4}$\\
C${}_{3}$H${}_{4}$ & $1.3 \times 10^{-10}$ & $1.0 \times 10^{3}$\\
CH${}_{3}$N & $1.9 \times 10^{-9}$ & $1.3 \times 10^{2}$\\
CH${}_{3}$OCH${}_{3}$ & $6.6 \times 10^{-8}$ & $6.0 \times 10^{1}$\\
CH${}_{5}$N & $1.2 \times 10^{-9}$ & $1.4 \times 10^{2}$\\
H${}_{2}$CO & $5.8 \times 10^{-8}$ & $2.2 \times 10^{4}$\\
HCOOCH${}_{3}$ & $3.1 \times 10^{-9}$ & $6.6 \times 10^{3}$\\
HNC & $1.1 \times 10^{-8}$ & $1.5 \times 10^{1}$\\
N${}_{2}$O & $8.5 \times 10^{-10}$ & $5.0 \times 10^{1}$\\
NH${}_{2}$CHO & $6.9 \times 10^{-8}$ & $2.8 \times 10^{1}$\\
NO & $6.1 \times 10^{-10}$ & $4.1 \times 10^{4}$ \\
\hline
\multicolumn{3}{l}{\textit{Comparing to initial point model}} \\
\hline
HNC & $1.1 \times 10^{-8}$ & $8.1 \times 10^{1}$\\
NO & $6.1 \times 10^{-10}$ & $1.5 \times 10^{3}$
\enddata
\end{deluxetable}

\subsection{Carbon and nitrogen in organics}
\label{sec:carbon-nitrogen}

One interesting question for planet formation is whether adding dynamics changes the overall organic composition at different disk locations\footnote{Here, ``organic'' refers to a gas- or solid-phase molecule or ion that contains both hydrogen and carbon.}. We assess this by considering the fraction of N and C in organic form in the static and dynamic tracks.

\begin{figure}[t]
	\includegraphics[width=\linewidth]{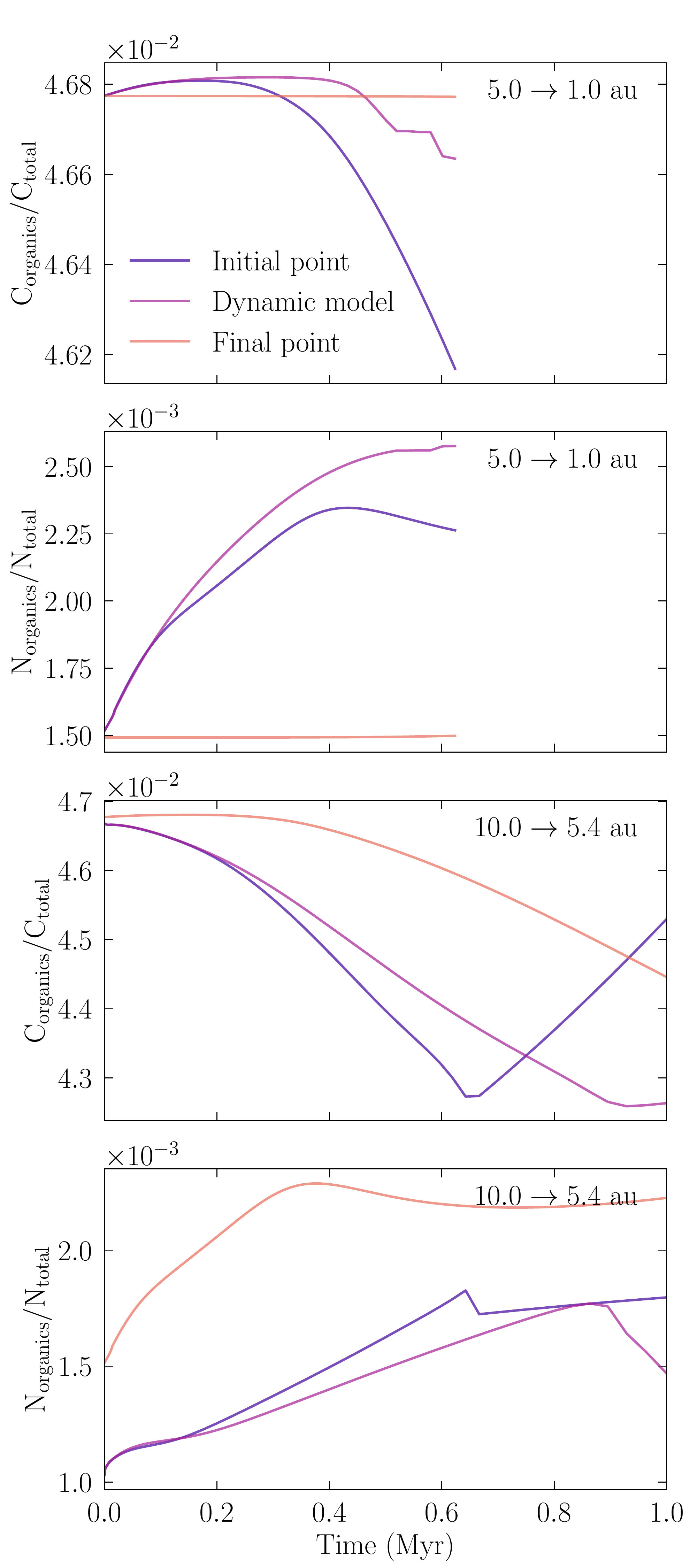}
    \caption{Time evolution of the amount of carbon and nitrogen in organics (gas and grain), expressed relative to the total amount of carbon and nitrogen, respectively. \firstrevision{Two different initial radii, 5~\protect\au\ and 10~\protect\au, are shown; the top two panels are for the 5~\protect\au\ track, while the bottom two panels are for the 10~\protect\au\ track.}}
    \label{fig:results3}
\end{figure}

In Figure~\ref{fig:results3}, we show the time evolution of the total amount of carbon and nitrogen in organics for the different dynamic tracks considered in the previous section compared to their static counterparts. The carbon in organics changes very slightly on the 5~\au\ dynamic track. The change increases moving to the 10~\au\ dynamic model, but is still modest; adding dynamics changes the carbon in organics by less than 10\%. We note that the effect of adding dynamics on the fraction of carbon in organics can produce different outcomes: The dynamic $5 \rightarrow 1$~\au\ model ends with an intermediate carbon fraction in organics compared to the two static counterparts, while the dynamic $10 \rightarrow 5.4$~\au\ model ends with a lower carbon fraction than either static model.

The fraction of nitrogen in organics is more sensitive to whether or not dynamics is taken into account. In the $5 \rightarrow 1$~\au\ dynamic model, the nitrogen fraction in organics ends at a value higher than either static model but is closer to that of the initial point model; in the $10 \rightarrow 5.4$~\au\ dynamic model, the nitrogen fraction is instead lower than either static model but is still closer to the value along the initial point model. There is an almost $50\%$ change in the nitrogen fraction along both the 5~\au\ and 10~\au\ dynamic tracks.

At the end of the 5~\au\ dynamic track, the nitrogen-bearing organic with the highest abundance is \ce{NH2CHO}. \ce{NH2CHO} forms efficiently from \ce{NH2} and \ce{H2CO} at early times, and this enhancement is then transported inwards, increasing the overall nitrogen fraction in organics. At the end of the 5~\au\ track, \ce{NH2CHO} comprises about 35\% of all nitrogen in organics.

At the end of the 10~\au\ dynamic track, the nitrogen-bearing organic with the highest abundance is instead \ce{H2CN} ice. This molecule accounts for about 52\% of all nitrogen in organics at the end of the track.

\subsection{Effect of cosmic rays}
\label{sec:cosmic-rays}

We may suspect that cosmic rays play a significant role in the disk chemistry of our dynamic models, given their stated importance in previous static models \citep{Cleeves+2014Science,Eistrup+2016A&A}. Evidence for this behavior comes from the fact that the dynamic track's evolution tends to more closely resemble the initial point model than the final point model, so much of the chemical processing must happen near the initial position of the gas parcel. Without cosmic rays playing a large role, this would be counterintuitive, since, for inward-moving tracks, the end of each track has higher temperatures and densities that would drive chemical processing at a higher rate than at the beginning of the track. By contrast, the cosmic ray flux in the midplane is at its \textit{highest} at the beginning of the track (Figure~\ref{fig:tracks}), where the surface density is lowest. Cosmic rays, then, provide a mechanism by which chemistry may be faster at lower disk temperatures and densities. 

\begin{figure*} %[h!]
	\includegraphics[width=\linewidth]{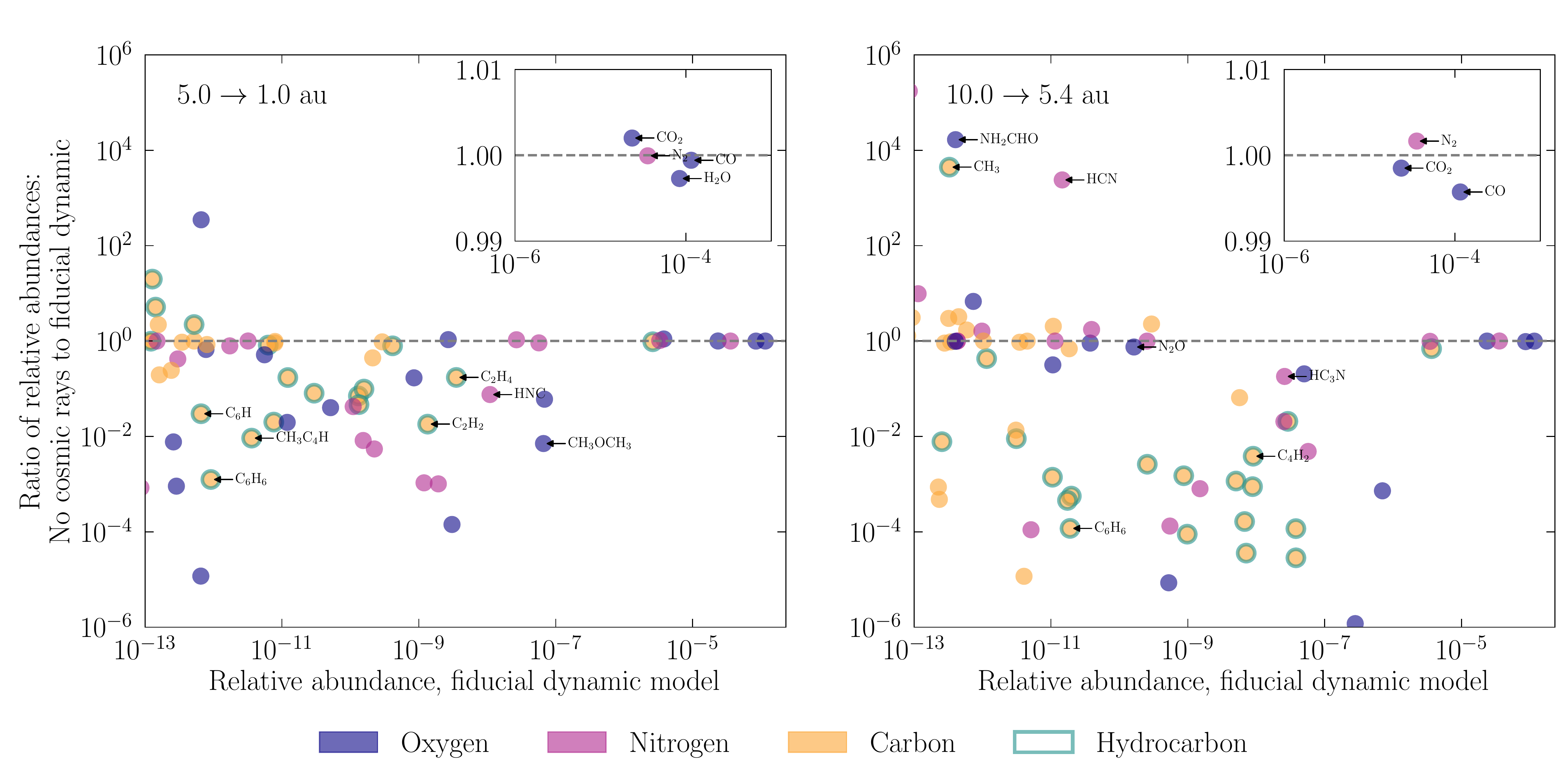}
	\caption{Comparison of two dynamic track models to identical models without cosmic rays. The gray dashed line indicates where the two models would produce the same results. Note that the value plotted is the total (gas + grain) relative abundance of each species. Inset axes are included to emphasize the small, but sometimes significant, enhancement or depletion of very abundant species, such as \ce{H2O} and \ce{CO2}.}
    \label{fig:compare-crays}
\end{figure*}

Figure~\ref{fig:compare-crays} shows a comparison between the $5\rightarrow1$ au and $10\rightarrow5$~au fiducial dynamic models and their counterparts with the cosmic ray flux set to zero at all times and radii. We clearly see that including cosmic rays has a dramatic effect on the chemistry along these tracks. Including cosmic rays  increases the abundances of many species. We note that this is especially true for the families of molecules that were enhanced in the fiducial models compared to he static models, such as hydrocarbons, which generally appear below the dashed line, where the two models would have equal abundances.

\section{Discussion}
\label{sec:discussion}

\subsection{General trends}

\begin{figure}%[ht]
    \includegraphics[width=\linewidth]{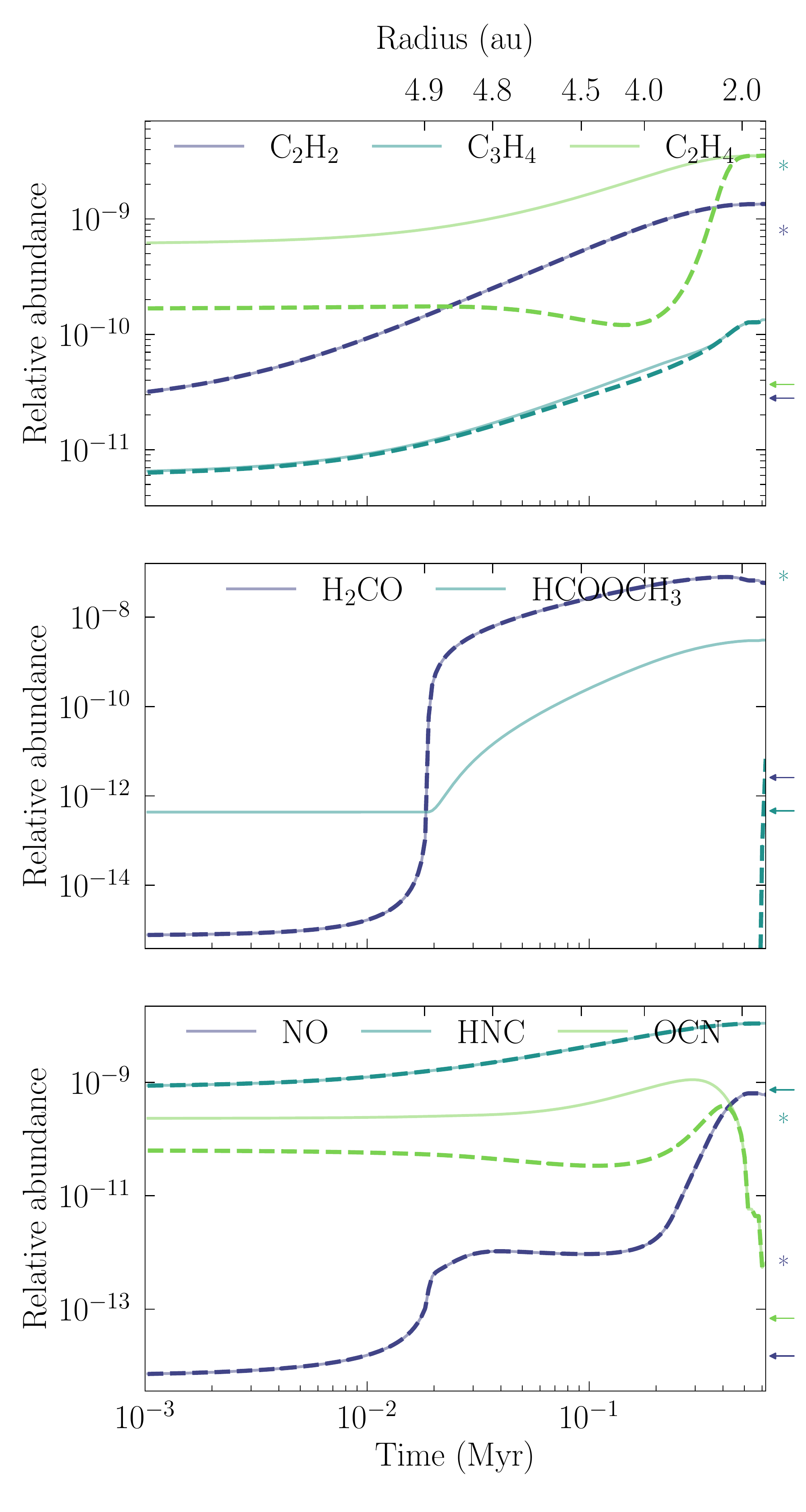}
	\caption{Time-dependent behavior of several species called out in the text, all along the 5~\au\ dynamic track and its static counterparts. Solid lines show the time-dependent behavior of the total abundance in our fiducial model, while dashed lines show the gas phase evolution in the fiducial model. At the far right, we show the final values corresponding to the initial point (stars) and final point (arrows) models.}
    \label{fig:4panel}
\end{figure}

In this section we discuss the origin of the enhancement patterns we see in the 5~\au\ and 10~\au\ tracks. %In both cases we find that cosmic rays play a major role in determining which species are enhanced and which are depleted. 
%The full, time-dependent behavior of each species mentioned in this section is shown in Figure~\ref{fig:4panel}.
In general, we find that including cosmic rays tends to enhance many species by the endpoint of an inward-moving track and we thus suspect that many of the observed trends can be traced back to a cosmic-ray driven chemistry close to the initial point of the tracks where cosmic ray penetration to the midplane is the most efficient. Figure~\ref{fig:4panel} (top and middle panels) shows that for the species enhanced in the dynamics $5\rightarrow1$~au track compared to the final point model, the chemical abundances are indeed mainly set during the first few hundred thousand years, when the gas parcel is $> 4$~\au.  % (see \S \ref{sec:cosmic-rays}). 

%Not every increase in a hydrocarbon species can be attributed to cosmic rays.
In more detail, we can see from Figure~\ref{fig:4panel} (top panel) that some of the most abundant and enhanced hydrocarbons in our 5~\au\ model --- namely, \ce{C2H2}, \ce{C3H4}, and \ce{C2H4} --- are produced rapidly at very early times and then experience a plateau until about 0.01~\Myr, when they experience a second rapid growth. The initial increase in \ce{C3H4} and \ce{C2H4} is simply because atomic carbon is present in our initial condition, and this atom reacts readily to form these products. We expect this to hold at all radii. 

The second growth step is more interesting and we examine the reaction rates, which allows us to isolate the dominant reaction pathways, for several species of interest. A representative example is \ce{C2H2}, a hydrocarbon species observed in the inner disk with \textit{Spitzer} (and may soon be observed with the \textit{James Webb Space Telescope}) that is enhanced at the end of the $5 \rightarrow 1$~\au\ track both when compared with the final point model and when compared to the endpoint of the same dynamic track without cosmic rays.  We find the following pathway for the formation of \ce{C2H2} at 0.05~\Myr\ along the 5~\au\ track.
\begin{equation}
    \begin{aligned}
        \ce{He &\myrightarrow{\mathrm{CRP}} He^+ + e^-} \\
        \ce{He^+ + CO &\myrightarrow{} C^+ + O + He} \\
        \ce{C^+ + CH4 &\myrightarrow{} C2H3^+ + H} \\
        \ce{C2H3^+ + NH3 &\myrightarrow{} NH4^+ + C2H2}
    \end{aligned}
    \label{eqn:c2h2-mechanism}
\end{equation}
Gas-phase \ce{C2H2} is thus produced by a reaction chain that begins with the ionization of helium. Through an electron exchange, a \ce{C^+} ion is produced, which then reacts with methane and ammonia to finally produce neutral \ce{C2H2}.

A similar analysis of  \ce{C3H4}, shows that it forms from \ce{C2H3+}, where \ce{C2H3+} is produced through the same pathway as listed above for \ce{C2H2}.
\begin{equation}
    \begin{aligned}
        \ce{C2H3+ + CH4 &\myrightarrow{} C3H5+ + H2} \\
        \ce{C3H5+ &\myrightarrow{\mathrm{gr}^-} C3H4 + H}
    \end{aligned}
\end{equation}

The mechanism for producing \ce{C2H4}, below, goes through a different set of species but still traces back to the high cosmic ray rate at 5~\au:
\begin{equation}
    \begin{aligned}
        \ce{C+ + CH3OH &\myrightarrow{} CH4O+ + C} \\
        \ce{C + H2 &\myrightarrow{} CH2} \\
        \ce{H + CH2 &\myrightarrow{} CH + H2} \\
        \ce{CH + CH4 &\myrightarrow{} C2H4 + H}.
    \end{aligned}
\end{equation}

%\red{basically HCOOCH3 forms from H2CO... so that's why they are linked? It seems like this is a lot of text for this idea.}

In summary, all hydrocarbons that are observed to be enhanced in the dynamic model, compared to the final point model, are enhanced due to the high level of cosmic ray ionization on the dynamic track compared to the final point model. 

In Figure~\ref{fig:results2}, we see that a few oxygen-bearing organics, including \ce{H2CO} and \ce{HCOOCH3}, are both enhanced along the 5~\au\ track relative to its static final counterpart; yet, the related species \ce{CH3OH} is not significantly enhanced or depleted, though it is also more abundant than the aforementioned species. 
Figure~\ref{fig:4panel} (middle panel) shows the complete time evolution of these two species. \ce{H2CO} and \ce{HCOOCH3} have the same overall behavior, wherein the molecule has a plateau at early times, followed by a rapid growth beginning around 0.02~\Myr. %and it is worth exploring why it occurs in both these species at almost the same time but not for \ce{CH3OH}. 
We determine the main formation pathways for both molecules at this time, similar to our analysis for hydrocarbons above. \ce{H2CO} is formed primarily from gas phase chemistry at 0.02~\Myr\ by
\begin{equation}\label{lab:h2co}
    \begin{aligned}
        \ce{O + CH3 &\myrightarrow{} H2CO + H}.
    \end{aligned}
\end{equation}
The radical precursor CH$_3$ comes in part from
\begin{equation}
    \begin{aligned}
       % \ce{H2 &\myrightarrow{\mathrm{CRP}} H2+ + e-} \\
       % \ce{H2+ + H2 &\myrightarrow{} H3+ + H} \\
        \ce{H3+ + CO &\myrightarrow{} HCO+ + H2} \\
        \ce{HCO+ + CH3OH &\myrightarrow{} CH5O+ + CO} \\
        \ce{CH5O+ + CH3OH &\myrightarrow{} CH3OCH4+ + H2O} \\
        \ce{CH3OCH4+ &\myrightarrow{\mathrm{gr}^-} CH3 + CH4 + O}.
    \end{aligned}
    \label{eqn:h2co}
\end{equation}
Essentially, \ce{CH3OH} is acting as a catalyst for the reactions but is not significantly produced or destroyed in the reaction scheme. Similar to the hydrocarbons, the reaction chain is initiate by cosmic ray chemistry, which is responsible for the formation of \ce{H3+}. \ce{HCOOCH3} mostly exists in its ice form, but it is initially formed in the gas phase (our model does not include a grain surface pathway) and then subsequently freezes onto grains. The gas-phase formation of \ce{HCOOCH3} is 
\begin{equation}
    \begin{aligned}
        \ce{CH5O+ + H2CO &\myrightarrow{} H5C2O2+ + H2} \\
        \ce{H5C2O2+ &\myrightarrow{\mathrm{gr}^-} HCOOCH3 + H};
    \end{aligned}
\end{equation}
These reactions are initiated by the  \ce{H2CO} as described in Equation~\ref{eqn:h2co}, hence their linked time evolution.
%The rest of the mechanism proceeds as above.
 
Figure~\ref{fig:results2} shows that there are a few species, particularly \ce{NO} and \ce{HNC}, that are enhanced in our fiducial model compared to \textit{both} static point models. This is interesting because it shows that the combination of transport and chemistry can result in excess production of some molecules in the disk midplane beyond any static model predictions. Figure~\ref{fig:4panel} shows the time-dependent behavior of \ce{NO}, \ce{HNC}, and \ce{OCN}, a possible precursor of \ce{NO}. At 0.3~\Myr, \ce{HNC} is primarily formed by:
\begin{equation}
    \begin{aligned}
        \ce{HCO+ + HCN &\myrightarrow{} HCNH+ + CO} \\
        \ce{HCNH+ + NH3 &\myrightarrow{} NH4+ + HNC}.
    \end{aligned}
\end{equation}
% the mechanism below; the production of \ce{HCO+} is as above
At the same time point, \ce{NO} is produced from
\begin{equation}
    \begin{aligned}
        \ce{HCO+ + NH2CHO &\myrightarrow{} NH2CH2O+ + CO} \\
        \ce{NH2CH2O+ &\myrightarrow{\mathrm{gr}^-} OCN + 2 H2} \\
        \ce{O + OCN &\myrightarrow{} NO + CO}.
    \end{aligned}
\end{equation}
% in general equations should always be written in sentence form in journals, including punctuation!!
The OCN precursor shows a similar growth behavior as NO at early times, but the two deviate dramatically as the parcel moves inward in the disk. In general we do not see a close connection between the abundances of different precursors and the final products, which implies that the production of NO and HNC, and by analogy many other molecules, are impossible to predict without running the full chemical code, including the relevant dynamics.
%\green{Again, we emphasize that this behavior would be impossible to predict \textit{a priori}; it is necessary to run the full simulation in order to know if a particular species is enhanced or depleted compared to a static model.}

%To conclude this section of the discussion, 
While we can trace enhancements of species many species in the dynamical model back to the high cosmic ray rate at the beginning of the 5~\au\ track, this is not true for every single species. In addition to the transport of cosmic-ray initiated chemistry at larger radii, we also see some species who enhancement is due to a complex interaction between transport and local chemistry. %as compared to the final point model, which has a very low cosmic ray rate at all times due to the high degree of shielding at small radii. 
Second, we find that there are some chemical families that are more sensitive to the addition of dynamics then others. Hydrocarbons as a family tend to be enhanced in the dynamic model compared to the static final point model, as are some oxygen-bearing organics and complex nitriles (Figure~\ref{fig:results2}). We emphasize, however, that it is virtually impossible to know \textit{a priori} which particular species will be enhanced due to inward transport and chemistry and which will barely be effected without actually running the complete model. The one exception to this rule may be the survival of initially very abundant, stable molecules, which in our models maintain close to their initial abundances at all investigated times and locations.

\subsection{Comparison to existing models}
 
When comparing our dynamic and static chemistry model outcomes to the most similar model in the literature by \citet{Heinzeller+2011ApJ}, we find both similarities and differences. Table~\ref{tbl:heinzeller} summarizes the this comparison. \ce{H2O} is barely affected by the inclusion of dynamics in both models. \ce{NH3} is also not strongly affected in either model, though what little effect there is goes in opposite directions. Both models predict some \ce{CH3OH} depletion in dynamic compared to static 1~\au\ models, but the magnitude of the depletion differs. The biggest difference is for \ce{C2H2}, however, where we find a large enhancement when including dynamics due to inward transport of cosmic ray-mediated chemistry, while \citeauthor{Heinzeller+2011ApJ} finds a depletion.

\begin{deluxetable*}{lccccc}[ht!]
\tablecaption{Comparison to \citet{Heinzeller+2011ApJ} model ACR; we list the ratio of the dynamic model value (either number density or column density) to its corresponding static model value. A value of unity indicates no change, while values less than or greater than unity indicate depletion or enhancement, respectively. \label{tbl:heinzeller}}
\tablehead{%
\colhead{Species} & \colhead{This work} & \colhead{$10 \times \zeta_\mathrm{CR}$} & \colhead{$2 \times T_\mathrm{gas}$} &  \colhead{$2 \times T_\mathrm{gas}$ and $10 \times \zeta_\mathrm{CR}$} & \colhead{\citet{Heinzeller+2011ApJ}\tablenotemark{a}} %
}
\startdata
\ce{H2O}   & $1$   & $1$   & $1$   & $1$   & $1$ \\
\ce{NH3}   & $1$   & $0.8$ & $1$   & $0.9$ & $1$ \\
\ce{CH3OH} & $0.9$ & $0.4$ & $0.9$ & $0.4$ & $0.03$ \\
\ce{C2H2}  & $50$  & $200$ & $5$   & $1$   & $0.008$
\enddata
\tablenotetext{a}{\citet{Heinzeller+2011ApJ} Table 3 lists the column densities for the species of interest. Since we do not have column densities for our midplane model, we warn the reader against comparing the table values directly.}
\end{deluxetable*}

An important difference between the two models, and therefore a potential source of the different model outcomes, is the treatment of cosmic rays. The \citeauthor{Heinzeller+2011ApJ} model computes its cosmic ray ionization rate from the density profile and dust opacity of \citet{Nomura+2007ApJ}. This model is different from ours and will therefore predict different levels of attenuation. More importantly, \citeauthor{Heinzeller+2011ApJ} adopts an unattenuated cosmic ray ionizaton of  $\zeta_\mathrm{CR} = 10^{-17}~\s[-1]$, whereas we have used $\zeta_\mathrm{CR} = 10^{-18}~\s[-1]$ in our models, following models by \citet{Cleeves+2014ApJ}. To test whether this explains the different model outcomes, we reran our dynamic and static models with an order of magnitude higher cosmic ray ionization rate (see figure in the Appendix).  While an enhanced cosmic ray ionization rate has a clear impact on the disk chemistry and changes the relative enhancements of many molecules in the dynamics vs. static models, the\ce{C2H2} enhancement seen in the fiducial model is preserved.  The mechanism which produces \ce{C2H2} at early times is the same as that given in Equation~\ref{eqn:c2h2-mechanism}. Different cosmic ray ionization rates alone does hence not explain the model differences.

A second difference between the two models is that the \citet{Heinzeller+2011ApJ} disk is warmer than the model presented here. To explore if the different temperature profiles can explain the observed chemical differences, we also ran models with an artificially boosted temperature profile, keeping the tracks the same\footnote{This is not, strictly speaking, a fully consistent approach, since the temperature profile also influences the tracks through the surface density evolution equation.}. We tested this warmer disk at both the fiducial and increased cosmic ray rate. The results of these trials are summarized in Table~\ref{tbl:heinzeller}. No combination of parameters results in a depletion of \ce{C2H2}, but in the warmer disk with high cosmic ray flux (i.e. the model that is most similar to \citealt{Heinzeller+2011ApJ}), we no longer produce a substantial \ce{C2H2} enhancement. In this model fast reactions consume \ce{C2H2} at the final time, incorporating \ce{C2H2} into larger molecules like \ce{C5H4N+} and \ce{C6H5+}. We note that this test implies that both our and their model results are sensitive to the precise disk structural model, which needs to be taken into account when directly interpreting disk chemistry results from observations. 

% The physical structure models are also likely different, since \citeauthor{Heinzeller+2011ApJ} assumes a constant accretion rate $\dot{M}$, while ours is allowed to vary, and an accretion parameter $\alpha = 10^{-2}$, which is ten times larger than ours. Finally, our star is actively evolving, so its luminosity and radius change with time, while the \citeauthor{Heinzeller+2011ApJ} model star has fixed luminosity and radius.

\subsection{Simplifying assumptions}

\firstrevision{To make the code computationally efficient, we have imposed a number of simplifying assumptions. One such assumption is that we do not presently consider vertical mixing in our disk model. Other studies have considered the impact of vertical mixing of gas and with solids on disk chemistry. \citet{Furuya+2013ApJ} found in their models that vertical mixing significantly decreased the column density of water ice in the disk. \citet{Kama+2016A&A} found a sequestration of carbon due to the vertical transport of carbon- and oxygen-bearing material from the disk surface to the midplane, where it freezes out onto grains. \citet{Ciesla+2012Science} found that mixing of grains enhanced their UV exposure during the disk lifetime, which facilitates the production of organics.}

\firstrevision{To evaluate the potential impact of treating the midplane in isolation, we follow \citet{SemenovWiebe2011ApJS} and compute the turbulent mixing timescale for the disk parameters we use. Under our assumptions, we find the temperature-dependent terms cancel, and the turbulent mixing timescale becomes a function of radius only,
\begin{equation}
    \tau_\mathrm{phys} = h^2 / D_\mathrm{turb} = \frac{\mathrm{Sc}}{\alpha \Omega},
\end{equation}
where $\mathrm{Sc}$ is the Schmidt number, which encodes the efficiency of turbulent diffusivity \citep{SemenovWiebe2011ApJS}; $h$ is the scale height of the disk; $D_\mathrm{turb}$ is the diffusion coefficient; $\alpha$ is the dimensionless viscosity parameter; and $\Omega$ is the orbital angular velocity. Substituting the relevant numbers, and evaluating this expression at 1~\au\ and 10~\au, we find timescales of $160 \; \mathrm{Sc}$~\yr\ and $5000 \; \mathrm{Sc}$~\yr, respectively. Note that the value of $\alpha$ we assume, $10^{-3}$, is informed by measurements from \citet{Flaherty+2018ApJ}, who measure low turbulence in the TW Hya disk.}

\firstrevision{Taking $\mathrm{Sc} = 1$ and $\mathrm{Sc} = 100$ as two possible values (the same values considered by \citealt{SemenovWiebe2011ApJS}), these timescales will always be shorter than the $\sim 10^6$~\yr\ timescale for surface chemistry (neglecting tunneling) at the disk midplane quoted from \citet{SemenovWiebe2011ApJS}, and if mixing is efficient we would therefore expect it to change grain surface compositions. We also consider how $\tau_\mathrm{phys}$ compares to the gas-phase processes in the disk. Ion-molecule chemical reactions have a typical timescale on the order of $10^0$ -- $10^1$~\yr\ \citep{SemenovWiebe2011ApJS}, which is short compared to mixing time scales at all relevant disk radii. Whether or not mixing could affect our results is thus a complex question, which depends on the relative importance of gas and grain surface chemical processes. We note that hydrocarbons, the species most affected by including dynamics in our model, are mainly gas phase chemistry products and we therefore expect this result to hold, while many of O-bearing organics, which form partially or wholly on grains, may be more sensitive to mixing. This is also in line with the findings of \citet{SemenovWiebe2011ApJS}.}

\firstrevision{Whether inner disk midplanes are subject to substantial vertical mixing is somewhat unclear, however. The few observational constraints on disk turbulence that exist are based on observations of gas in the outer disk, and typically well above the midplane. Based on such observations, \citet{Teague+2016A&A}, for example, measured $v_\mathrm{turb} \sim 0.2$ -- $0.4 c_s$ in TW Hya. \citet{Hughes+2011ApJ} found $\alpha \sim 0.01$ in HD 163296, and \citet{Flaherty+2018ApJ} found evidence for low turbulence in TW Hya with $\alpha < 0.007$. These low turbulence measurements may not be surprising because the magneto-rotational instability (MRI) may not be as active as originally thought \citep{Simon+2018ApJ}. Additional observations are clearly needed to establish levels of turbulence at all disk scales as the effects will likely be chemically important.} %This does not preclude disk chemical studies of the effects of disk turbulence, however. Indeed, observations of ``disk midplane chemistry'' at disk surfaces may be one window onto disk mixing efficiencies, and we plan to extend our model framework to include vertical mixing in future publications.}

\firstrevision{In addition, the present prescription does not allow for mixing of the gas or mixing of the dust between different radial regions. Dust actively evolves by growth and fragmentation in protoplanetary disks \citep[e.g.,][]{DullemondDominik2005A&A}. These processes influence the dust surface area relative to volume, and therefore we expect it to impact chemistry. However, dust evolution likely cannot be explained by a simple monotonic growth, and therefore would require a full treatment of dust evolution, which is beyond the scope of this paper.}  
%. For example, the lifetime of an individual small grain is relatively short (on the order of one thousand years to collide and grow) as compared to the lifetime of the disk, but fragmentation can replenish the small grain population \citep{DullemondDominik2005A&A}. The larger grain evolution is further complicated by the radial drift problem \citep{}. The chemical code does not follow individual grains, and thus to address these issues will require an entirely new computational approach, which will be explored in our future work.}

%\subsection{Dust Evolution}

\section{Conclusions}
\label{sec:conclusions}

We have undertaken a self-consistent model of midplane disk chemistry and dynamical evolution that includes viscous accretion, under the assumption of well-coupled gas and dust. We find that taking accretion into account, and the associated changes in physical conditions along a gas parcel's journey, can substantially change the abundances of many species within 10~\au. Many of these species are enhanced because of cosmic ray-driven reactions in the outer disk, which are then transported into the ``cosmic ray dark'' inner disk regions where the gas attenuation is very high. There are, however, also species that are depleted when including dynamics, and predicting \textit{a priori} how the chemistry will be affect by the inclusion of dynamics  is challenging. Abundant species --- most notably \ce{H2O}, \ce{CO2} and \ce{CO} --- are largely unaffected by the inclusion of dynamics, so a static model would approximate their abundances well. %Any individual species' enhancement or depletion can be very sensitive to the incident flux of cosmic rays, and quantitative comparisons with observations will require better constraints on this fundamental parameter.

\secondrevision{Inner disk chemistry is much more strongly affected than outer disk chemistry, and the radii impacted are similar to those observed with \textit{Spitzer} and that will be observed with \textit{JWST}. \textit{Spitzer} has detected several molecules in protoplanetary disks, including \ce{H2O}, \ce{OH}, \ce{C2H2}, \ce{HCN}, and \ce{CO2} \citep{Salyk+2008ApJ,Pontoppidan+2010ApJ}, and we expect \textit{JWST} to make many more detections. Are the hydrocarbons seen by \textit{Spitzer}, such as \ce{C2H2}, native to the disk atmosphere, or were they lofted up from the midplane by vertical mixing? The answer to this question depends on the strength of vertical mixing, and so constraining its nature warrants further observational study.} %better constraints so we cannot immediately make predictions for observations without more work on the magnitude of turbulence in disks.}

\acknowledgements

E.M.P. gratefully acknowledges support from National Science Foundation Graduate Research Fellowship Program (GRFP) grants DGE1144152 and DGE1745303. This work was supported by an award from the Simons Foundation (SCOL \# 321183, KO).

\software{RADMC-3D \citep{radmc3d}, Computation of isochrones \citep{Siess+2000A&A}}

\bibliographystyle{aasjournal}
\bibliography{biblio,petsc-surfdens,petsc-chemphysdisk}

\appendix

\begin{figure*}[hp]
    \centering
    \includegraphics[width=\linewidth]{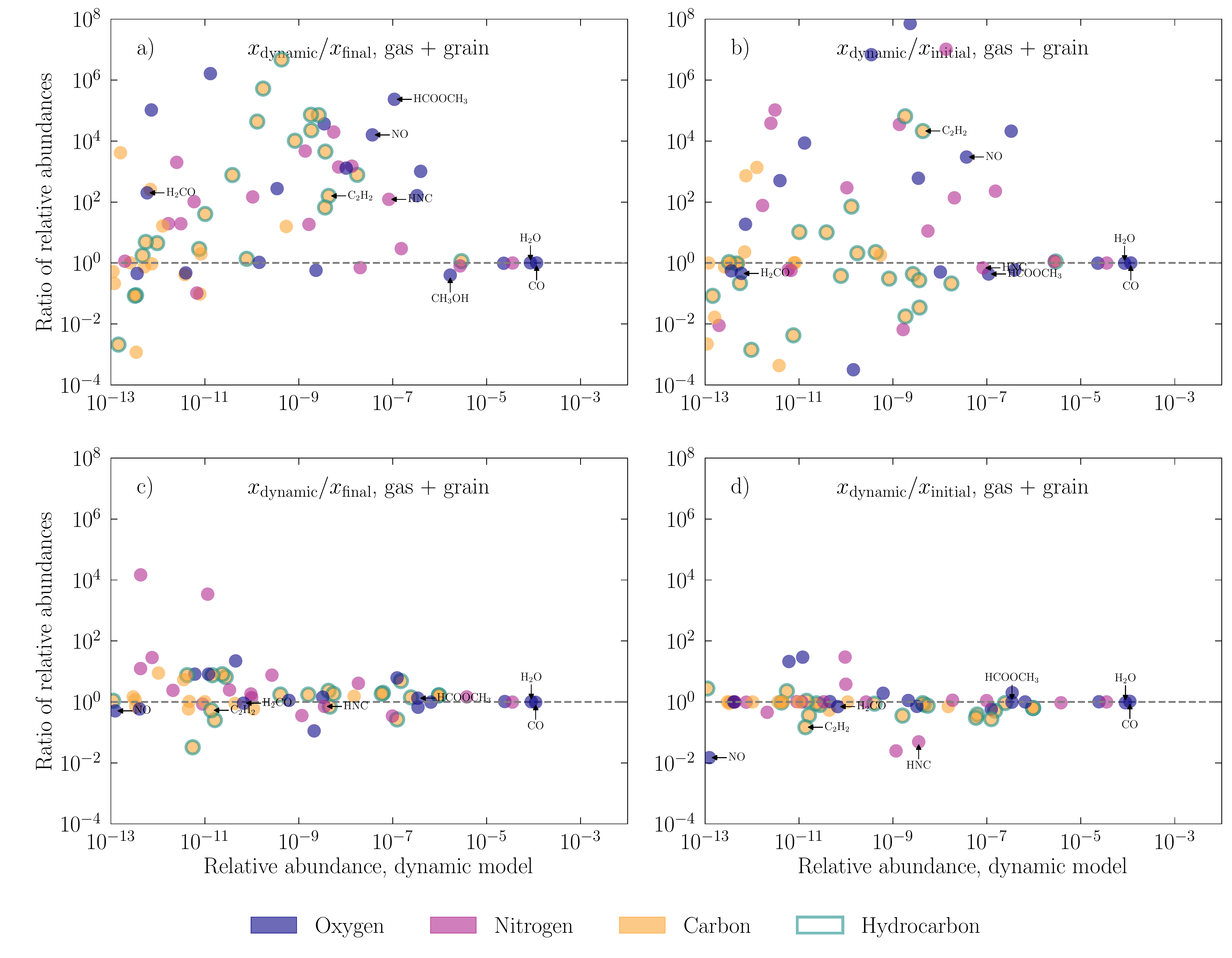}
    \caption{Analogue of Figure~\ref{fig:results2} using the higher cosmic ray ionization rate consistent with \citet{Heinzeller+2011ApJ}, $\zeta_\mathrm{CR} = 10^{-17}~\protect\s[-1]$.}
    \label{fig:results2_hicrays}
\end{figure*}
\end{document}